\documentclass[aps,physrev,preprint,groupedaddress]{revtex4-2}

\usepackage{graphicx} 
\usepackage{siunitx} 
\usepackage{tabularx}
\usepackage{booktabs}
\usepackage{multirow}
\usepackage{amsmath}

\usepackage{hyperref}
\hypersetup{
    colorlinks,
    citecolor={blue},
    urlcolor={blue}
}

\begin{document}


\title{Assessing AI in Introductory Physics Problem Solving}


\author{Amir Bralin}
\email[Contact email: ]{abralin@ttu.edu}
\affiliation{Texas Tech University, Lubbock, TX}

\author{N. Sanjay Rebello}
\affiliation{Purdue University, West Lafayette, IN}


\date{\today}

\begin{abstract}
Reasoning or inference-scaling models are the new generation of Large Language Models (LLMs) capable of complex problem solving. 
To investigate their problem-solving capability in physics, we evaluated model \mbox{o4-mini} by \mbox{OpenAI} on solving traditional, end-of-chapter problems from Halliday and Resnick's ``Fundamentals of Physics,'' spanning core topics in the undergraduate physics curriculum. 
Performance was analyzed across modality and problem difficulty. 
The model solved the problems with overall accuracy of about 90\%, but performance depended strongly on representation: accuracy was much higher on text-only problems (96\%) than on problems requiring coordinated interpretation of text and images (79\%). 
Accuracy also declined significantly as the problem difficulty increased from low to medium to high. 
These results show that state-of-the-art LLMs can solve much of the standard introductory physics problems, but that their performance remains uneven and constrained by problem modality and problem difficulty.
\end{abstract}


\maketitle

\section{Introduction}
The term Artificial Intelligence (AI) refers to one of the following: (1) ``the capability of computer systems or algorithms to imitate intelligent human behavior''; (2) ``a computer, computer system, or set of algorithms having this capability''; or (3) ``a branch of computer science dealing with the simulation of intelligent human behavior by computers''~\cite{mw:ai}.
To preserve semantic rigor in this work, we refer to ``AI'' in sense 1 only.
Sense 2 is referred to as ``AI models'' or ``AI systems'' interchangeably.
Sense 3 is referred to as ``AI research.''
Recent advancements in AI research, particularly with the release of \mbox{GPT-4} in 2023~\cite{OpenAI2023}, have considerably expanded the capabilities of AI models in general and their applications in physics.
These models can process both textual and visual inputs, enabling researchers to evaluate their performance across a wide range of physics tasks.
At the current state of affairs, AI models demonstrate strong text-based problem-solving abilities, but continues to struggle with visual interpretation, complex reasoning, and reliable execution of multi-step solutions.

Studies in Physics Education Research (PER) evaluated AI on standardized physics assessments~\cite{Polverini2024a, Polverini2025a}.
Performance on FCI~\cite{Hestenes1992, Hake1998} showed highly inconsistent results: some questions are answered correctly every time, while others are never solved correctly~\cite{Aldazharova2024}.
On the TUG-K~\cite{Beichner1994, Zavala2017}, where models perform significantly worse on image-based tasks than on the text-based ones, this issue was reinforced~\cite{Polverini2024b}.
Across multiple state-of-the-art AI models, difficulty with extracting meaning from graphs (such as the axis labels or the slope) remains a central weakness~\cite{Polverini2025b}.
In addition to visual limitations, AI struggles with under-specified physics problems that require assumptions or estimation.
When faced with real-world problems lacking complete numerical information, models frequently fail to infer reasonable physical conditions.
And even when problems are fully specified, errors still occur due to incorrect physical modeling or simple calculation mistakes~\cite{Wang2024}.

Despite these shortcomings, AI consistently demonstrates strength in identifying relevant physical concepts, suggesting that its conceptual understanding is more robust than its procedural execution.
The lack of self-evaluation after attempting to solve a problem is especially pronounced in the default model behavior~\cite{Kieser2024}.
Targeted prompting strategies (such asking the model to justify and double-check its own answer) can mitigate this issue, indicating that model performance is sensitive to how problems are \textit{framed}.
Another important factor influencing AI performance is problem \textit{complexity}.
While models perform well on simpler scenarios involving straightforward applications of physics principles, their accuracy drops significantly when problems involve more complex phenomena~\cite{Dunlap2025}.
This suggests that AI models ``struggle'' to integrate multiple layers of physical reasoning, particularly when additional conceptual or mathematical complexity is introduced.

As newer models are developed, researchers expand evaluation studies to include more advanced and diverse physics problems: Olympiad problems~\cite{Tschisgale2025}, multilingual datasets~\cite{Kortemeyer2025b}, and context-rich problems~\cite{Horchani2025}.
Across an extensive array of problem types in physics education, the state of the art in AI continues to improve and approach human-level performance (in some cases, such as in text-based problem solving---even the expert level).
One intriguing study showcased that AI models are capable of solving newly released problems ensuring that they cannot rely on prior exposure during training~\cite{Yu2025}.
Additional benchmarking studies in AI research provide a broader perspective by testing models on thousands of physics problems across multiple domains, including graduate-level material~\cite{Chung2025}.
Qualitative analyses revealed two key issues: first, models often fail to incorporate implicit real-world knowledge unless explicitly stated; second, they frequently make errors in complex mathematical reasoning, particularly in multi-step symbolic calculations~\cite{Feng2025, Zheng2025}.

\subsection{Research Questions}
Overall, the literature presents AI models as a powerful but yet imperfect tool for physics problem solving. 
Its strengths lie in simulating conceptual understanding, pattern recognition, and adaptability across problem types. 
Processing visual input, simulating implicit human knowledge, and executing complex calculations limit its effectiveness.
As AI continues to improve, ongoing evaluation allows to track its progress and ensure its appropriate use in educational and scientific contexts.
To this end, we addressed the following questions:
\begin{enumerate}
    \item What is the problem-solving capability of AI models on standard topics in the introductory physics curriculum?
    \item How does this capability depend on modality---language and vision?
    \item How does this capability depend on different levels of difficulty in the problems solved?
\end{enumerate}
An additional property of a model when solving analytical problems is the amount of ``effort'' it exerts while engaging in this process.
This quantity may be defined as the number of \textit{tokens} produced by the model in generating the full solution to a given problem.
It provides auxiliary information about the model behavior.
From it, one may also calculate the financial cost of running the model.

\section{Methods}
Among various AI models made available at the time of data collection and analysis in this study, ``\mbox{o4-mini}'' by \mbox{OpenAI} was selected due to its design and affordability.
Together with a larger model ``o3,'' they were designed ``to think for longer before responding''~\cite{OpenAI2025a}, allowing for better performance on problem-solving tasks compared to their predecessors such as ``\mbox{GPT-4o}'' and ``o1.''
At the same time, \mbox{o4-mini} is almost twice cheaper than \mbox{o3}, and thus is more affordable to everyday users.
It was released at about the same time~\cite{OpenAI2025b}.
We used \mbox{OpenAI} Application Programming Interface (API) at \url{https://openai.com/api/} to access both models.
All Python programming and calls to \mbox{OpenAI} API were done in Google Colab (\url{https://colab.google/}).

The newest version of \mbox{GPT-5} was published soon afterwards.
However, OpenAI no longer refers to it as a ``model'' but rather as a ``system'' that includes a faster ``efficient model that answers most questions,'' a slower ``reasoning model {\dots} for harder problems,'' and a ``router that quickly decides which to use''~\cite{OpenAI2025c}.
Thus, before evaluating the capability of this sophisticated system, we decided to report the results for an earlier, simpler model that is \mbox{o4-mini}.

Among various physics textbooks that contain end-of-chapter problems, ``Fundamentals of Physics'' by \citeauthor{Walker2022}~\cite{Walker2022} was chosen due to its status and popularity in the undergraduate physics curriculum.
It consists of two volumes: Vol.~1 covers standard topics in mechanics, thermodynamics, and kinetic theory;
Vol.~2 covers standard topics in electromagnetism, optics, relativity, and introductory quantum mechanics.
There are problems at the end of each chapter in the textbook.
Each problem is given a difficulty level by the authors: Easy, Medium, and Hard.
Answers are provided for only odd-numbered problems.
In this study, we assume that the textbook's answer key contains no errors and may serve as ground truth in our analysis.

All odd-numbered problems along with their answers from the textbook PDF were manually copied and pasted into a blank {\LaTeX} document.
All English words and numbers (in arabic numerals) were preserved during this process.
However, equations (math) and special symbols (Greek letters) had to be re-typed manually.
Some units of measurement as well as numbers under the scientific notation had to also be re-typed case by case.
We used package \mbox{siunitx}~\cite{Wright2025} to cast all such instances under a unifying, typographical notation.
We believe that such consistent approach aligns well with the well-structured, text processing method of language models.

\begin{figure}
    \centering
    \includegraphics[width=\linewidth]{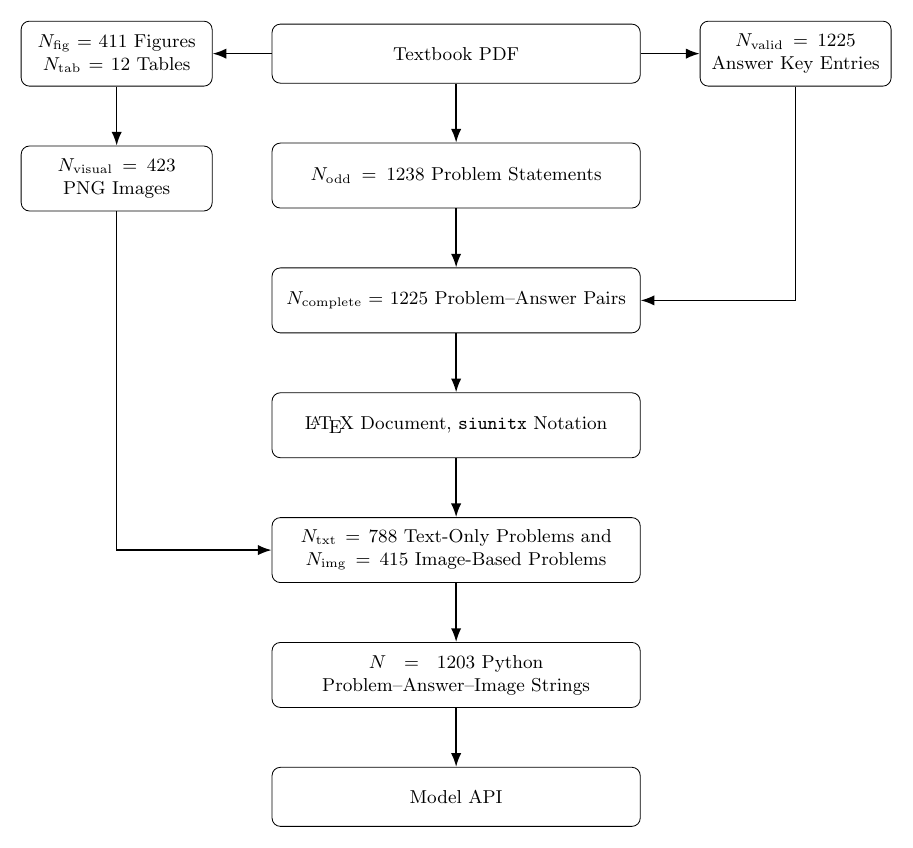}
    \caption{Process of converting textbook end-of-chapter problems into AI model input}
    \label{fig:data_flowchart}
\end{figure}

Some problems involved images, most of which were figures and only a few of which were tables.
We extracted all relevant images by simply making screenshots.
To preserve the relative resolution of the images with respect to each other, all screenshots were made at the fixed size of the displayed PDF pages.
As a result, 411 images were obtained and saved under the PNG format, with the sizes ranging from 6 to 610 kB.
Still some problems, while not having any images associated with them at first, asked to show or sketch some visual representation (diagram, graph).
They did not have any answers provided in the textbook, and thus, were discarded from our data.
In the end, we had $N = \num{1203}$ problems in the dataset.
Figure~\ref{fig:data_flowchart} shows each step of this process as a flowchart.

Table~\ref{tab:vol1} shows the contents of the textbook's first volume.
Table~\ref{tab:vol2} shows the contents of the second volume.
Modality indicates whether the problems were only in the form of text only `TXT' or in the combined form (text + image) `IMG.'
Difficulty indicates the level (`E' for easy, `M' for medium, and `H' for hard) of difficulty provided by the textbook authors.
As is shown on the tables, there was about the same number of problems in each volume, with twice as many of text-only problems as there were combined.
In addition, there were much fewer hard-level problems than there were the easy and medium ones.
In terms of physics topics, the textbook reflects the standard undergraduate curriculum.
Chapters 1--17 cover mechanics (with chapters 15--17 being on oscillations and waves).
Chapters 18--20 cover thermodynamics (with chapter 19 being on the kinetic theory).
Chapters 21--36 cover electrodynamics (with chapters 21--32 being on electromagnetism and chapters 34--36 being on optics).
Chapter 37 is about the theory of relativity and chapters 38--40 serve as an introduction to the quantum theory.

All problem-answer pairs thus formatted in {\LaTeX}, along with any corresponding images, were transferred to a spreadsheet for each textbook chapter.
The problem numbers as given in the textbook as well as their difficulty levels were also transferred to the spreadsheet.
So, each row in the spreadsheet corresponded to a distinct problem and contained all information needed to solve it.
Using standard packages and libraries in Python (such as \verb|numpy| and \verb|pandas|), we iterated through each row in the spreadsheet and provided model \mbox{o4-mini} with the problem text and image(s) (if any).
We did not provide it with the number and difficulty level of any problem.
So, every run of the model's problem solving was \textit{independent}.

Once all solution runs for all problems from a given chapter were completed, we instructed a powerful model \mbox{GPT-5}~\cite{OpenAI2025c} with scoring them as either correct or incorrect according to the textbook answers.
That is, similar to \mbox{o4-mini}, this model was given as input: the problem text and images (if any), the answer, and the previously generated solution.
This was done for every run per problem, resulting in $N \times n = \num{6015}$ distinct scores obtained by the model.
To verify these results, we evaluated a sample of $m = 600$ generated solutions and evaluated the scores assigned by \mbox{GPT-5}.
Out of 600, only 14 scores were assigned incorrectly: that is \num{97.67}\%.
Assuming that the textbook answer key contains no errors, we set the \textit{systematic error} in such AI-based scoring to be fixed and equal to \num{0.023}.
Together with the statistical error, it makes up the total uncertainty in model accuracy achieved over all problems considered in the dataset.

To verify the relationship between the model's accuracy and the problem difficulty, we performed three statistical tests.
The first was a simple logistic regression, with difficulty levels treated as an ordinal variable.
This assumed that all individual solution runs are \textit{independent} and yields overall statistics about the relationship.
In addition, it also assumed the differences in-between two levels are the same: Medium problems are just as more difficult than Easy problems, as Hard problems are more difficult than the Medium ones.
The second statistical test accounted for repeated runs per individual problem, which is more appropriate since the model is likely to behave in a similar way when encountering the same exact problem albeit for a new run.
The third test released the assumption about \textit{uniform} difficulty and, instead, treated both Medium and Hard categories separately, in relation to level Easy (base).
With that, we established the final, most reliable fit between accuracy and difficulty and reported the associated statistics.
Python library \verb|statsmodels| was used to perform all tests.
For more details, refer to Appendix~\ref{ap:stat_testing}

\begin{table*}[tbhp]
    \footnotesize
    \caption{\label{tab:vol1}Vol.~1 chapter titles and the corresponding numbers of problems.}
    \centering
    \begin{tabular}{ l  c  c  c  c  c }
        \textbf{Chapter} & \multicolumn{5}{c}{\textbf{Problems}}\\
        \hline
         & \multicolumn{2}{c}{\textbf{Modality}} & \multicolumn{3}{c}{\textbf{Difficulty Level}} \\
        \hline
         & \textbf{TXT} & \textbf{IMG} & \textbf{E} & \textbf{M} & \textbf{H} \\
        \hline
        \phantom{0}1. Measurement                                   & 13 &  2 &  8 &  5 & 2 \\
        \phantom{0}2. Motion Along a Straight Line                  & 25 & 10 & 15 & 16 & 4 \\
        \phantom{0}3. Vectors                                       & 17 &  5 & 13 &  9 &  \\
        \phantom{0}4. Motion in Two and Three Dimensions            & 33 &  8 & 13 & 22 & 6 \\
        \phantom{0}5. Force and Motion--I                           & 18 & 16 & 11 & 20 & 3 \\
        \hline
        \phantom{0}6. Force and Motion--II                          & 15 & 14 & 10 & 16 & 3 \\
        \phantom{0}7. Kinetic Energy and Work                       & 17 &  9 &  9 & 15 & 2 \\
        \phantom{0}8. Potential Energy and Conservation of Energy   & 12 & 18 &  9 & 15 & 6 \\
        \phantom{0}9. Center of Mass and Linear Momentum            & 22 & 18 & 13 & 23 & 4 \\
        10. Rotation                                                & 25 &  9 & 14 & 16 & 4 \\
        \hline
        11. Rolling, Torque, and Angular Momentum                   & 18 & 17 & 16 & 14 & 5 \\
        12. Equilibrium and Elasticity                              &  6 & 18 & 10 & 11 & 3 \\
        13. Gravitation                                             & 26 &  8 & 18 & 13 & 3 \\
        14. Fluids                                                  & 26 &  9 & 18 & 17 &  \\
        15. Oscillations                                            & 20 & 12 & 17 & 12 & 3 \\
        \hline
        16. Waves--I                                                & 23 &  7 & 13 & 15 & 2 \\
        17. Waves--II                                               & 29 &  6 & 16 & 15 & 4 \\
        18. Temperature, Heat, and the First Law of Thermodynamics  & 23 & 10 & 14 & 17 & 2 \\
        19. The Kinetic Theory of Gases                             & 26 &  6 & 13 & 17 & 2 \\
        20. Entropy and the Second Law of Thermodynamics            & 16 &  6 &  8 & 11 & 3 \\
        \hline
        & \textbf{410} & \textbf{208} & \textbf{258} & \textbf{299} & \textbf{61} \\
    \end{tabular}
\end{table*}

\begin{table*}[tbhp]
    \footnotesize
    \caption{\label{tab:vol2}Vol.~2 chapter titles and the corresponding numbers of problems.}
    \centering
    \begin{tabular}{ l  c  c  c  c  c }
        \textbf{Chapter} & \multicolumn{5}{c}{\textbf{Problems}}\\
        \hline
         & \multicolumn{2}{c}{\textbf{Modality}} & \multicolumn{3}{c}{\textbf{Difficulty Level}} \\
        \hline
         & \textbf{TXT} & \textbf{IMG} & \textbf{E} & \textbf{M} & \textbf{H} \\
        \hline
        21. Coulomb's Law                                       & 11 &  8 &  6 & 10 & 3 \\
        22. Electric Fields                                     & 15 & 13 & 10 & 17 & 1 \\
        23. Gauss' Law                                          & 16 & 12 & 12 & 13 & 3 \\
        24. Electric Potential                                  & 22 & 12 & 12 & 19 & 3 \\
        25. Capacitance                                         & 15 &  7 & 11 & 16 & 1 \\
        \hline
        26. Current and Resistance                              & 24 &  3 & 10 & 16 & 1 \\
        27. Circuits                                            & 13 & 20 & 12 & 18 & 3 \\
        28. Magnetic Fields                                     & 23 &  9 & 17 & 14 & 1 \\
        29. Magnetic Fields Due to Currents                     & 12 & 20 & 13 & 16 & 3 \\
        30. Induction and Inductance                            & 19 & 19 & 18 & 17 & 3 \\
        \hline
        31. Electromagnetic Oscillations and Alternating Current& 20 & 10 & 15 & 15 &  \\
        32. Maxwell's Equations; Magnetism of Matter            & 16 &  9 &  9 & 15 & 1 \\
        33. Electromagnetic Waves                               & 19 & 14 & 15 & 17 & 1 \\
        34. Images                                              & 12 &  6 &  8 &  9 & 1 \\
        35. Interference                                        & 13 &  8 & 12 & 18 & 1 \\
        \hline
        36. Diffraction                                         & 29 &  6 & 21 & 14 &  \\
        37. Relativity                                          & 26 &  3 & 11 & 17 & 1 \\
        38. Photons and Matter Waves                            & 33 &  2 & 10 & 23 & 2 \\
        39. More About Matter Waves                             & 17 &  6 &  9 & 14 &  \\
        40. All About Atoms                                     & 25 &  4 & 18 & 11 &  \\
        \hline
        & \textbf{380} & \textbf{207} & \textbf{249} & \textbf{309} & \textbf{29}\\
    \end{tabular}
\end{table*}

\subsection{Metrics}
Let's denote the score obtained by model \mbox{o4-mini} as a binary quantity:
\begin{equation}
    s = \begin{cases}
            1 \quad \text{if correct}\\
            0 \quad \text{otherwise}
        \end{cases}
\end{equation}
For a set of $N$ problems the model obtains a set of scores $\{ s_i \}$, where $i = 1, 2, \dots, N$.
The \textit{accuracy} of the model is defined as the number of problems it solved correctly.
For the binary case, it is calculated by the sum of all scores:
\begin{equation}
    \text{accuracy} = \frac{1}{N}\sum_{i = 1}^{N}{\{s_i\}}.
\end{equation}
The scores on each problem were assigned based on the final answer to that problem.

To obtain statistical results, we repeated the process of solving each problem $n = 5$ times.
That is, the model generated its solution to a given problem repeatedly.
This is important because language models are statistical, with some degree of variability in their responses to an identical prompt.
When their output is large in size, with many operations involved, text generated at the end may easily deviate from one instance to another.
The model may successfully solve a given problem at one instance, while obtaining wrong results at another.
To obtain the average score for every problem, we averaged over all runs:
\begin{equation}
    \text{problem score} = \frac{1}{n}\sum_{i = 1}^{n}{\{s_i\}},
\end{equation}
where $s_i$ is the score of the $i$-th run.
The associated \textit{statistical error} was calculated by the standard error of the mean (SEM) $\sigma_{\text{prob}}/\sqrt{n}$, where $\sigma_{\text{prob}}$ is the standard deviation of the problem accuracy.

To obtain the average accuracy for every chapter, we averaged over all $N_i$ problems in a given chapter $i$:
\begin{equation}
    \text{chapter accuracy} = \frac{1}{N_i}\sum_{j = 1}^{N_i}{\{\text{problem score}_j\}}, \quad i=1, 2, \dots, 40
\end{equation}
The statistical error associated with problem variation within each chapter was also calculated by SEM $\sigma_{\text{chap}}/\sqrt{N_i}$.
The additional errors coming from run variation within each chapter problem were propagated, adding to the total statistical error in our chapter accuracy results.
Finally, to report the overall accuracy for all textbook chapters considered together, we averaged over all problems in the textbook \textit{without dividing them into chapters}.
To calculate the associated statistical error, we followed the same procedure just described.

\section{Results}
\begin{figure}[htb]
    \centering
    \includegraphics[width=\linewidth]{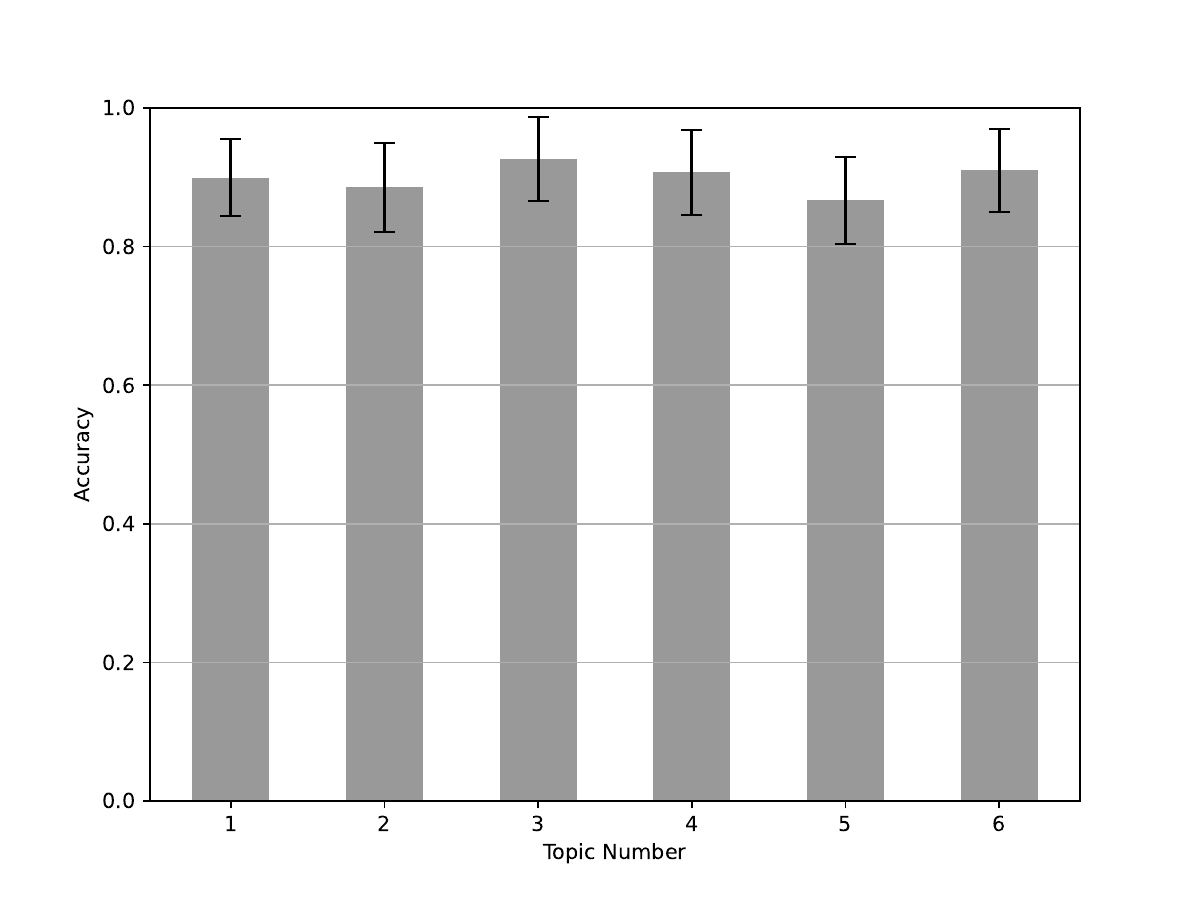}
    \caption{Accuracy results by physics topics.}
    \label{fig:topic_scores}
\end{figure}

Tables \ref{tab:scores_vol1} and \ref{tab:scores_vol2} show the scores achieved by model \mbox{o4-mini} on all problems from 40 chapters from \citeauthor{Walker2022} as graded by model \mbox{GPT-5}.
The column Text indicates the model accuracy on text-only problems, while column Text + Image indicates the accuracy on problems with images when the model had to process both modalities simultaneously.
Evidently, for such hybrid problems the model performed much worse when compared with the text-only ones.
The average accuracy achieved on all problems from each chapter overall is shown at the bottom row of each table.
It is similar for Vol.~1 and Vol.~2, indicating the uniform distribution of the model's problem solving capability across topics broadly.
For better visualization of the uniformity of scores across chapters, we aggregated them by six topics shown on Table~\ref{tab:topic_scores} and plotted a bar chart in Figure~\ref{fig:topic_scores}.

\begin{figure}[htb]
    \centering
    \includegraphics[width=0.8\linewidth]{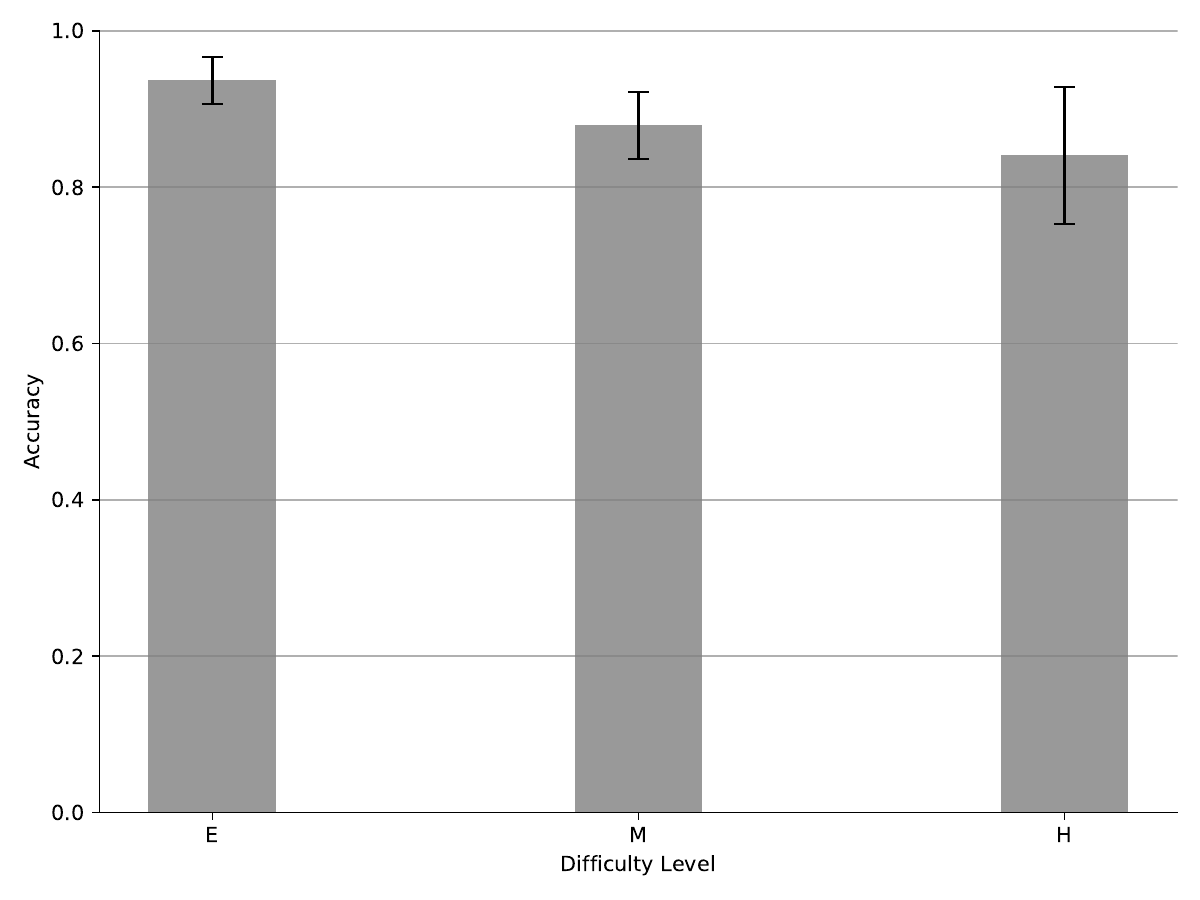}
    \caption{Accuracy results by problem difficulty.}
    \label{fig:level_scores}
\end{figure}

Overall, the model scored $0.90 \pm 0.03$ in accuracy (out of 1.0).
We can see the drastic difference between the accuracy of the model on text-only problems ($0.96 \pm 0.03$) and on problems with images ($0.79 \pm 0.04$).
Table~\ref{tab:level_scores} shows aggregated accuracy scores grouped by three levels of difficulty (Easy, Medium, Hard) in the problems solved.
We can see the gradual drop in accuracy corresponding to a jump in difficulty: first by about \num{0.06} from Easy to Medium, and then by \num{0.04} from Medium to Hard.
Note that the uncertainty considerably increased for level Hard since there were far less problems in that category than there were in the categories Easy and Medium.
Figure~\ref{fig:level_scores} shows the distribution of model scores across the three levels.
Visually, the decrease is evident while the overall accuracy stays consistently above 0.8, which is quite impressive.

\begin{figure}[htb]
    \centering
    \includegraphics[width=0.8\linewidth]{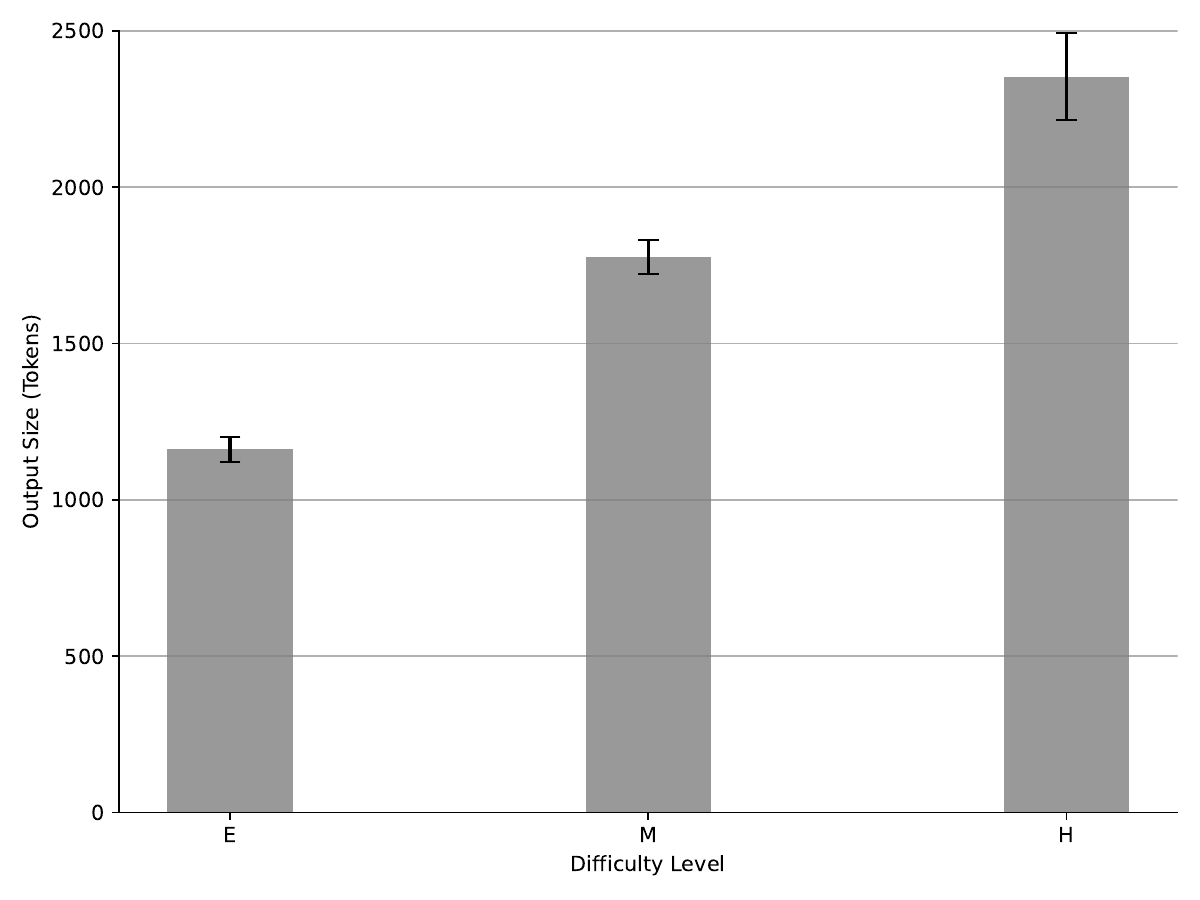}
    \caption{Average output tokens by problem difficulty.}
    \label{fig:level_tokens}
\end{figure}

Figure~\ref{fig:level_tokens} shows the distribution of model effort (as measured by output tokens) across three levels of difficulty in the problems solved.
As we can see, while accuracy gradually decreases, the output size generated by the model to keep up with rising difficulty decreases correspondingly.
In total, \num{10.7e6} tokens were used by the model: \num{1.3e6} for processing input and \num{9.4e6} for generating output.
This represents the computational cost as typically one token corresponds to a single (floating-point) operation in the computer.
Knowing the rates at which \mbox{OpenAI} charges per million tokens, it is straightforward to calculate the financial cost too.

\begin{table*}[htb]
\centering
\caption{Vol.~1 scores by chapter and modality. Some chapter titles were abbreviated for visual purposes. (CoE: Conservation of Energy, AM: Angular Momentum, FLT: First Law of Thermodynamics, SLT: Second Law of Thermodynamics)}
\label{tab:scores_vol1}
\begin{tabular}{l c c c}
\toprule
\textbf{Chapter} & \textbf{Text} & \textbf{Text + Image} & \textbf{Overall} \\
\midrule
1. Measurement                              & 0.97(0.07) & 1.00(0.00) & 0.97(0.07) \\
2. Motion Along a Straight Line             & 1.00(0.00) & 0.46(0.25) & 0.85(0.10) \\
3. Vectors                                  & 0.98(0.06) & 0.48(0.32) & 0.86(0.12) \\
4. Motion in Two and Three Dimensions       & 0.98(0.05) & 0.75(0.19) & 0.94(0.07) \\
5. Force and Motion--I                      & 0.94(0.08) & 0.82(0.17) & 0.89(0.11) \\
6. Force and Motion--II                     & 1.00(0.00) & 0.94(0.10) & 0.97(0.06) \\
7. Kinetic Energy and Work                  & 0.99(0.05) & 0.78(0.20) & 0.92(0.09) \\
8. Potential Energy and CoE                 & 1.00(0.00) & 0.84(0.16) & 0.91(0.11) \\
9. Center of Mass and Linear Momentum       & 1.00(0.00) & 0.82(0.14) & 0.92(0.08) \\
10. Rotation                                & 0.94(0.08) & 0.71(0.20) & 0.88(0.09) \\
11. Rolling, Torque, and AM                 & 0.91(0.11) & 0.75(0.15) & 0.83(0.11) \\
12. Equilibrium and Elasticity              & 0.93(0.13) & 0.59(0.23) & 0.68(0.19) \\
13. Gravitation                             & 0.99(0.04) & 1.00(0.00) & 0.99(0.03) \\
14. Fluids                                  & 0.98(0.05) & 0.87(0.16) & 0.95(0.07) \\
15. Oscillations                            & 0.98(0.06) & 0.97(0.08) & 0.98(0.06) \\
16. Waves--I                                & 0.93(0.09) & 0.17(0.19) & 0.74(0.12) \\
17. Waves--II                               & 0.90(0.11) & 1.00(0.00) & 0.92(0.10) \\
18. Temperature, Heat, and the FLT          & 0.97(0.06) & 0.84(0.18) & 0.93(0.08) \\
19. The Kinetic Theory of Gases             & 0.93(0.10) & 0.83(0.19) & 0.91(0.10) \\
20. Entropy and the SLT                     & 0.94(0.09) & 0.93(0.13) & 0.94(0.09) \\
\midrule
\hfill \textbf{Average}:                             & \textbf{0.96(0.05)} & \textbf{0.79(0.10)} & \textbf{0.90(0.06)} \\
\bottomrule
\end{tabular}
\end{table*}

\begin{table*}[htb]
\centering
\caption{Vol.~2 scores by chapter and modality. Some chapter titles were abbreviated for visual purposes. (AC: Alternating Current, MoM: Magnetism of Matter)}
\label{tab:scores_vol2}
\begin{tabular}{l c c c}
\toprule
\textbf{Chapter} & \textbf{Text} & \textbf{Text + Image} & \textbf{Overall} \\
\midrule
21. Coulomb's Law                               & 0.96(0.08) & 0.95(0.11) & 0.96(0.08) \\
22. Electric Fields                             & 0.95(0.09) & 0.82(0.20) & 0.89(0.13) \\
23. Gauss' Law                                  & 0.94(0.10) & 0.80(0.17) & 0.88(0.11) \\
24. Electric Potential                          & 0.97(0.07) & 0.92(0.11) & 0.95(0.07) \\
25. Capacitance                                 & 1.00(0.00) & 0.78(0.20) & 0.90(0.11) \\
26. Current and Resistance                      & 1.00(0.00) & 1.00(0.00) & 1.00(0.00) \\
27. Circuits                                    & 1.00(0.00) & 0.88(0.12) & 0.93(0.08) \\
28. Magnetic Fields                             & 0.94(0.09) & 0.87(0.17) & 0.92(0.10) \\
29. Magnetic Fields Due to Currents             & 0.97(0.08) & 0.58(0.22) & 0.72(0.17) \\
30. Induction and Inductance                    & 0.95(0.08) & 0.89(0.12) & 0.92(0.08) \\
31. Electromagnetic Oscillations and AC         & 1.00(0.00) & 0.96(0.09) & 0.99(0.04) \\
32. Maxwell's Equations; MoM                    & 0.90(0.11) & 0.76(0.22) & 0.85(0.12) \\
33. Electromagnetic Waves                       & 1.00(0.00) & 0.73(0.20) & 0.88(0.10) \\
34. Images                                      & 1.00(0.00) & 0.67(0.27) & 0.89(0.11) \\
35. Interference                                & 0.97(0.07) & 0.86(0.13) & 0.90(0.10) \\
36. Diffraction                                 & 0.90(0.10) & 0.33(0.23) & 0.81(0.11) \\
37. Relativity                                  & 0.95(0.07) & 1.00(0.00) & 0.96(0.06) \\
38. Photons and Matter Waves                    & 0.84(0.12) & 0.80(0.22) & 0.84(0.13) \\
39. More About Matter Waves                     & 0.98(0.06) & 0.83(0.19) & 0.94(0.08) \\
40. All About Atoms                             & 0.93(0.09) & 0.90(0.18) & 0.92(0.09) \\
\midrule
\hfill \textbf{Average}: & \textbf{0.95(0.05)} & \textbf{0.81(0.10)} & \textbf{0.90(0.06)} \\
\bottomrule
\end{tabular}
\end{table*}

\begin{table}[htb]
    \centering
    \caption{Accuracy results by physics topics.}
    \label{tab:topic_scores}
    \begin{tabular}{c l r r l}
    \toprule
    \# & \textbf{Topic} & \textbf{Chapters} & $N_{\text{problems}}$ & \textbf{Accuracy} \\
    \midrule
         1 & Mechanics                          &  1--14 & 434 & 0.90(0.06) \\
         2 & Oscillations and Waves             & 15--17 &  96 & 0.89(0.06) \\
         3 & Thermodynamics and Kinetic Theory  & 18--20 &  87 & 0.93(0.06) \\
         4 & Electromagnetism                   & 21--32 & 353 & 0.91(0.06) \\
         5 & Electromagnetic Waves and Optics   & 33--36 & 117 & 0.87(0.06) \\
         6 & Relativity and Quantum Theory      & 37--40 & 116 & 0.91(0.06) \\
    \end{tabular}
\end{table}

\begin{table}[htb]
    \centering
    \caption{Accuracy results by problem difficulty.}
    \label{tab:level_scores}
    \begin{tabular}{c r l}
    \toprule
    \textbf{Difficulty Level} & $N_{\text{problems}}$ & \textbf{Accuracy} \\
    \midrule
         Easy   & 507 & 0.94(0.05) \\
         Medium & 607 & 0.88(0.07) \\
         Hard   &  89 & 0.84(0.11) \\
    \end{tabular}
\end{table}

\begin{table}[htb]
    \caption{
    Results of the first two statistical tests performed on accuracy vs difficulty data.
    The $z$-statistic is for coefficient $\beta_1$.
    }
    \label{tab:stat_test12}
    \centering
    \begin{tabular}{c c c c c}
         \toprule
         Test & $\beta_0$ & $\beta_1$ & $z$ & $P > \vert z \vert$ \\
         \midrule
         1 & 3.186(0.133) & -0.566(0.069) & -8.2 & 0.000 \\
         2 & 3.186(0.236) & -0.566(0.121) & -4.7 & 0.000 \\
    \end{tabular}
\end{table}

\begin{table}[htb]
    \caption{
    Results of the third statistical test performed on accuracy vs difficulty data.
    The 95\% confidence interval (CI) is for the odds-ratio (OR).
    }
    \label{tab:stat_test3}
    \centering
    \begin{tabular}{c c c c c c c}
         \toprule
         Coefficient & Mean & St.D. & $z$ & $P > \vert z \vert$ & OR & 95\% CI \\
         \midrule
         $\beta_0$ & 2.6909 & 0.152 & 17.693 & 0.000 & & \\
         $\beta_1$ & -0.7072 & 0.186 & -3.802 & 0.000 & 0.49 & (0.34, 0.71) \\
         $\beta_2$ & -1.0294 & 0.275 & -3.747 & 0.000 & 0.36 & (0.21, 0.61) \\
    \end{tabular}
\end{table}

\clearpage
\section{Conclusion and Implications}
From the results of this study, we found that the statistical language model \mbox{o4-mini} provided by \mbox{OpenAI}~\cite{OpenAI2025a} is capable of solving physics problems in a standard undergraduate curriculum, as represented by the \citeauthor{Walker2022}'s textbook~\cite{Walker2022}, with high accuracy $0.90 \pm 0.03$.
However, this largely depends on modality: the accuracy achieved for solving problems with images ($0.79 \pm 0.04$) is much lower than that for text-only problems ($0.96 \pm 0.03$).
There was no distinct pattern in topic-based model performance observed, as the accuracy scores stayed stable across chapters on average.
In contrast, clear trends were observed for the accuracy scores and the model effort (output tokens) based on problem difficulty.
With each jump in difficulty, the former decreased and the latter increased steadily.
The accuracy trend was confirmed by a series of statistical tests performed on it using standard techniques such as logistic regression and generalized estimating equations~\cite{Liang1986}.
Namely, the probability of solving a Medium-level problem is smaller than the probability of solving an Easy-level problem by 0.51 on average, and the probability of solving a Hard-level problem is, in turn, smaller than the probability of solving a Medium-level problem by 0.27 on average.
Finally, the amount of computational effort spent by the model on solving Medium and Hard problems almost doubled compared with that spent on the Easy ones.

The result that modality dictates model performance was not surprising, as it is well-established in the present literature~\cite{Polverini2025b}.
Building high-performance models equally capable of processing text and images remains a challenge for AI research.
This is recognized in the AI research community as well as in the AI industry.
NVIDIA, for instance, is pushing the frontier with its Physical AI paradigm of building hybrid models capable of handling visual and linguistic data equally well.
For the physics community this result implies that it is still going to take somewhat longer for AI models to become capable of complete problem-solver experience.
Since in physics, the inability to visualize effectively, to draw diagrams, to plot insightful graphs, and so on, is tantamount to not being able to solve great many problems at all.
Nevertheless, given the pace of progress it seems to be only a matter of time.

The result that the model was able to solve problems from different physics topics, ranging from classical mechanics to quantum mechanics, uniformly on average also aligns with some of the results obtained previously by \citeauthor{Polverini2024b}~\cite{Polverini2024b} for earlier \mbox{OpenAI} models.
They tested the model on kinematics graphs and concluded that ``it appears that grouping the items in terms of test objectives does not provide novel meaningful insights into the strengths and weaknesses of individual chatbots'' (where test objectives were referred to as the ``areas of conceptual understanding for which the test was developed'').
That is, they didn't observe any clear patterns in model performance based on different test items corresponding to variable content knowledge.
For example, one item involved acceleration while another involved velocity---the two concepts are distinct and novice students don't always grasp the two equally well at first (see, for example, early studies in PER \cite{Trowbridge1980, Trowbridge1981}).
Nevertheless, for the AI model that didn't matter: what it ``sees'' is two similar graphs with vectors on each and processes both either equally well or equally bad.
From a practical point of view, this result is reassuring for physics teachers and students who use such state-of-the-art models in their work or study.
It implies a fair amount of reliability and freedom from topic-associated deficiencies on part of the model.

The result that the accuracy of the model decreased significantly with rising difficulty in the problems attempted to solve, albeit common-sense, is a relatively novel finding for the PER community and even for broader AI research community.
Previous PER studies that evaluated AI models on certain tests with various problem types (\cite{Dunlap2025, Tschisgale2025, Kortemeyer2025b, Polverini2025b}) did not include any similar results.
Related studies in AI research (\cite{Zheng2025, Yu2025}) did report an observed decline in model accuracy associated with different levels of problems involved.
However, no detailed statistical analysis was performed and no rigorous pattern was established.
It also has broader implications.
If we assume that the model behaves consistently and each solution attempt that it runs on a given problem, then this may serve as a measure of problem difficulty.
Such measure would be determined computationally and, therefore, more \textit{objective}.
In our case, for example, we may suggest the following interpretation of this result: the difficulty levels assigned by the textbook authors follow an increasing order (from Easy to Medium to High), but it is not uniform.
And hypothetically, when designing a test bank and trying to classify a problem set based on difficulty, one could use an AI model to do that objectively.

\subsection{Limitations and Future Work}
In the context of physics, and especially for educational purposes, AI models must exhibit valid reasoning and problem solving.
A number of assessment tools for measuring these characteristics (along with concept inventories and epistemological surveys) were developed by the PER community~\cite{Adams2015, Docktor2016, Price2021}.
Originally designed for assessing student learning, they are readily applicable to AI as well.
Thus applied, they can show how closely AI models simulate student thinking.
If the results indicate that AI exhibits expert-like conceptual understanding, problem-solving skills, or attitudes and beliefs, then it would mean that these models have achieved an expert-like level of performance in the domain of physics.

Evaluating problem-solving capability of such AI models is important for making decisions about using them in problem solving.
Students may want use AI technology for their learning.
In a setting where the answer to a physics homework problem is known but the solution is not, a hypothetical student may resort to AI-generated solution for assistance.
As reported by \citeauthor{Elby2025}~\cite{Elby2025}, they may do so in order ``to check their reasoning'' or ``to get themselves `unstuck' when they couldn't figure out a step in the problem-solving process.''
As long as there are no images involved, this strategy may prove useful for student learning.
However, AI-generated solutions to standard physics problems need to be evaluated for following sound problem-solving methods.
Understanding how these models achieve the correct answer or by what means is important for ensuring their safety and reliability.

Teachers and educators may want to use AI for assessment.
This is especially relevant to schools with large enrollment classes.
Language models in particular offer a powerful tool for assessing qualitative skills, often hidden from quantitative testing.
For instance, students may be instructed to not only solve a given physics problem but also to explain their solution in writing.
Our work may serve as only a prerequisite for implementing this strategy.
It is one thing for an AI model to correctly solve a problem, it is another thing for it to evaluate another solution based on that.

\appendix
\section{Prompting}
\label{ap:prompting}
This section contains the snippets of Python code we wrote for prompting \mbox{OpenAI} models through its API.
The user prompt that model o4-mini received for solving every problem in our dataset is shown below.
\begin{verbatim}
model_response = client.responses.create(
  model="o4-mini-2025-04-16",
  reasoning={"effort": "medium"},
  input=[
      {
        "role": "user",
        "content": input_content,
      }
  ],
  max_output_tokens=20000,
)
\end{verbatim}
The model checkpoint \verb|o4-mini-2025-04-16| was specified, although no further changes have been introduced since the model's release date.
The parameter \verb|reasoning_effort| controls the amount of ``effort'' (tokens) that the model exerts on ``reasoning'' (generating text hidden from the user) while processing given input.
By default, it is set to \verb|medium| and we kept it as such.
We played the ``role'' of the ``user'' and provided \verb|input_content| to the model which consisted of the problem statement (in text) and related images (if any).
Finally, we set the maximum number of tokens generated at output to \num{20000}.
No problem solution in all of our iterated runs has reached this limit.

The code below shows how we constructed the input content of each prompt for o4-mini.
\begin{verbatim}
input_content = [{"type": "input_text", "text": problem_text}]

for problem_image in problem_images:
  image_path = root + problem_image
  base64_image = encode_image(image_path)
  input_content.append({"type": "input_image",
                        "image_url": f"data:image/png;base64,{base64_image}"
                        })
\end{verbatim}
For the problems that didn't have any images (figures and tables), the input content was simply set to be the problem text.
No further instructions such as ``You need to solve this problem...'' or ``You are a physics expert...'' were added.
For the problems with images, each related image was attached to the input content under Base64 encoding.
Some problems involved two or three images.

\section{Statistical Testing}
\label{ap:stat_testing}
We can describe the binary score $s$ achieved by the model (the binary outcome of whether the model correctly solves a problem) on a given problem as a function of difficulty.
Let $s_{ij}$ denote the correctness of the $j$-th run on problem $i$:
\begin{equation}
    s_{ij} = \begin{cases}
            1 \quad \text{if correct,}\\
            0 \quad \text{otherwise.}
        \end{cases}
\end{equation}
Using logistic function we can write:
\begin{equation}
\text{logit}\big(P(s_{ij} = 1)\big) = \beta_0 + \beta_1 \cdot \text{difficulty}_i,
\end{equation}
where $\text{logit}(p) = \log\left(\frac{p}{1-p}\right)$ and difficulty level encoded in uniform order:
\begin{equation}
    \text{difficulty}_{i} = \begin{cases}
                            1 \quad \text{if Easy}\\
                            2 \quad \text{if Medium}\\
                            3 \quad \text{if Hard}\\
                            \end{cases}
\end{equation}
Then, the difficulty acts as the predictor for our binary label which is the model score.
Here the key assumption is that the scores change \textit{linearly} with difficulty and \textit{at the same rate}.
That is, the relative change from Easy to Medium is assumed to be the same as that from Medium to Hard.
Estimating the coefficients of this relation using the maximum likelihood estimator, we can find out the statistical effects of increasing difficulty on the overall accuracy scores.
This translates into a hypothesis test:
\begin{equation}
\begin{cases}
    H_0: \beta_1 = 0 \quad \text{then no effect} \\
    H_1: \beta_1 \neq 0 \quad \text{then difficulty affects accuracy}
\end{cases}
\end{equation}
For a large sample as in our dataset, the Wald test is used to estimate the coefficients.

To account for repeated runs of the same problem, we can deploy logistic regression with generalized estimating equations~\cite{Liang1986}, grouping the scores of runs corresponding to the same problem together.
This approach adjusts for within-problem correlation and provides more reliable standard error estimation.
To account for non-uniform changes in difficulty among levels, we can add the requirement to treat $\text{difficulty}_i$ ($i = 1, \dots, N$) as a categorical variable, encoded in a triplet like a vector.
Then, both Medium and Hard categories are offset against the category Easy, which acts a base.
In the logistic function, it translates into fitting $\beta_0 + \beta_1 \cdot \text{Medium} + \beta_2 \cdot \text{Hard}$.

Table \ref{tab:stat_test12} shows the test results with and without the assumption about independent runs.
Test 1 fits simple logistic regression to the accuracy vs difficulty data and yields negligible $P$-value ($< 0.05$), indicating the linear (on the log-odds scale) relationship between the two.
Test 2 fits a generalized linear model to the same data but with individual solution runs grouped together under the same problem.
It also yields negligible $P$-value, confirming the previous test results.
And it even yields the same mean coefficients.
The difference between the two test results is in the standard deviation of each coefficient: each value from Test 2 is about twice larger than that from Test 1.
This is appropriate because accounting for sample runs per a problem must result in a more reliable measure of uncertainty.
That is, a larger one.

Table \ref{tab:stat_test3} shows the test results assuming that increases in difficulty are not identical.
The same $z$-statistic was applied to each coefficient, resulting in negligible $P$-values again.
The odds ratio (OR) indicates the effect size of each level: each one-level increase in difficulty is associated with a OR\% reduction in the odds of a correct response.
That is, the likelihood of correctness drops by $1 - \text{OR}$ for each level jump.
So, the odds-ratio of 0.49 for Medium level means the 51\% reduction in the odds of model \mbox{o4-mini} solving the problem compared to Easy level.
Similarly, the OR of 0.36 for Hard means the 64\% reduction in that likelihood.
To find the OR value for $\text{Medium} \rightarrow \text{Hard}$ jump, we simply divide the two since ORs are multiplicative:
\begin{equation}
    \text{OR}_{M \rightarrow H} = \frac{\text{OR}_{E \rightarrow H}}{\text{OR}_{E \rightarrow M}} = 0.73.
\end{equation}
Comparing these effect sizes across levels reveals rather a non-linear pattern: the largest decrease in performance occurs between Easy and Medium levels (51\%), with a smaller additional decrease from Medium to Hard (27\%).
This suggests that the impact of increasing difficulty is not uniform across levels, but instead exhibits diminishing marginal effects at higher difficulty.

\begin{acknowledgments}
This research was supported in part by NSF Grant 2300645.
Any opinions, results, and findings expressed here are those of the authors and not of the NSF.
\end{acknowledgments}

\bibliography{ref.bib}
\end{document}